# Cooperative Anti-Jamming for UAV Networks: A Local Altruistic Game Approach

SU Yueyue[a], QI Nan[a,*], HUANG Zanqi[a], YAO Rugui[b], JIA Luliang[c],

[a] *College of Electronic Information Engineering, Nanjing University of Aeronautics and Astronautics, Nanjing 210016, China*
[b] *School of Electronics and Information, Northwestern Polytechnical University, Xian 710072, China*
[c] *School of Space Information, Space Engineering University, Beijing 101416, China.*

## Abstract

To improve the anti-jamming ability of the UAV-aided communication systems, this paper investigates the channel selection optimization problem in face of both internal mutual interference and external malicious jamming. A cooperative anti-jamming method based on local altruistic is proposed to optimize UAVs' channel selection. Specifically, a Stackelberg game is modeled to formulate the confrontation relationship between UAVs and the jammer. A local altruistic game is modeled with each UAV considering the utilities of both itself and other UAVs. A distributed cooperative anti-jamming algorithm is proposed to obtain the Stackelberg equilibrium. Finally, the convergence of the proposed algorithm and the impact of the transmission power on the system loss value are analyzed, and the anti-jamming performance of the proposed algorithm can be improved by around 64% compared with the existing algorithms.

*Corresponding author. Tel.: +86 18351406807.
E-mail address: nanqi.commun@gmail.com

## 1. Introduction

Since the low-altitude unmanned aerial vehicle (UAV) can establish short-range line-of-sight (LoS) links with the ground units, and is faster to deploy, it is often used to assist information dissemination and collection. As one of modern high-tech equipment, UAVs are widely adopted in agriculture, firefighting, and military departments[1-3] with admirable performance. However, the UAV is limited by payload during mission, poor flight capability and insufficient data processing equipment. With these inevitable shortcomings, a single UAV communication device is difficult to cope with the challenges of complex tasks and a harsh environment. Multiple UAVs are more capable of processing tasks than a single UAV. Therefore, it usually takes a group of UAVs to accomplish a mission.

The communication system of UAVs is vulnerable to malicious jammers. Especially when the UAVs perform tasks in clusters, there is not only external malicious jamming but also serious internal interference in the environment. Recent works related to optimizing the anti-jamming performance of the UAV communication system mainly focus on three areas: power control, spectrum allocation and path planning.[4-6] Specifically, in Ref. 4, the incomplete information and the co-channel mutual interference were considered, and a Bayesian Stackelberg game was modeled to

This work was supported in part by the National Natural Science Foundation of China under Key Project (No. 61931011), the National Natural Science Foundation of China (No. 61827801, 61801218, 61901523, 62071223), the Natural Science Foundation of Jiangsu Province (No. BK20180424), the Fundamental Research Funds for the Central Universities, NO. NT2021017 formulate the relation between UAVs and the jammer. There is no information exchange among users in Ref. 5, and a channel selection algorithm was proposed to improve the throughput and interference mitigation performance. Moreover, a path planning method of jammer evasion was proposed to guarantee communication quality in Ref. 6. But the cooperative anti-jamming and coordination capabilities among UAVs are not considered in such literatures, which limits the application in multi-user systems.

With the growth of jamming attacker's capabilities, the traditional individual anti-jamming methods are difficult to deal with intelligent complex and strong jamming. Moreover, due to the limited resources that a single UAV can be equipped with, its anti-jamming performance is often difficult to meet the real needs. However, the multi-UAV intelligent cooperative anti-jamming method is superior to traditional methods.[7] The difference between traditional anti-jamming methods and cooperative ones is that the former means the legitimate nodes only care about their own utilities and separately occupy spectrum resources. But the latter mean resources sharing, not "selfish", and cares about the interests of themselves and other nodes simultaneously. When the legal communication unit is subjected to malicious jamming from the jammer, it can avoid interference by adjusting its own channel selection. And when mutual interference occurs due to communication requirements, multiple UAVs can choose different communication channels to improve reliability and effectiveness.

In recent years, several literatures about cooperative anti-jamming have been proposed to address the traditional shortcomings. The existing cooperative anti-jamming methods mainly include four categories: cooperative transmission,[8-9] cooperative anti-jamming strategy,[10-11] cooperative information interaction[12-13] and cooperative system framework.[14] Cooperative transmission is one of the most common cooperative anti-jamming methods. In Ref. 8, a collaborative UFH-based (CUB) broadcast scheme was proposed to improve the communication efficiency against the smart jammer. And it was performed over universal software radio peripherals in Ref. 9. The CUB allows nodes to relay received packets to other nodes so that the communication efficiency and anti-jamming performance can be improved. Cooperative anti-jamming strategy refers to the system reducing the interference by using honeypots[10] or other ways. An idea named "No Pains No Gains" was designed to trap the jammer by sacrificing parts of users' benefits in Ref. 11. Cooperative information interaction means interactive cooperation among users and improves the anti-jamming capability of the system. A collaborative multi-agent anti-jamming algorithm was proposed in Ref. 12, and users realized coordination by Q values exchange to obtain the optimal anti-jamming strategy. In Ref. 13, users interact with the neighbor node to determine the communication payoff, according to which the channel access probability is updated. Cooperative system framework uses the different perceptions, computing abilities and communication resources among communication nodes to achieve intelligent collaboration. In Ref. 14, cooperative cognitive radio network can switch among three kinds of architectures autonomously and flexibly, which can deal with electromagnetic interference. It can be observed that all these cooperative anti-jamming works focus on the jamming from malicious jammers and co-channel users to improve the defense performance. However, users in these studies pay little attention to the local utility of the system. It is a fundamental and important problem to study the optimal channel selection under the local impact of users.

To tackle the above problems, this paper considers the scenario where the UAVs perform telemetry tasks and send information to the fusion center (FC) along different routes. UAVs are affected by both malicious intelligent interference and co-channel interference of all other UAVs. Inspired by Ref. 15, each UAV considers the payoffs of itself and all other UAVs. UAVs choose different channels to induce the interference with each other. To obtain the optimal channel selection of UAVs, a distributed cooperative anti-jamming channel selection algorithm is proposed based on stochastic learning automata (SLA). UAVs and the jammer can learn from the utility function and adjust their channel selection strategy flexibly.

To summarize, the major contributions of this paper are given as follows:

1) The confrontation relationship between UAVs and the jammer is modeled as a Stackelberg game, where the jammer is the leader and the UAVs are the followers. The cooperative anti-jamming problem among UAVs is modeled as a local altruistic game. The sum payoff of itself and all other UAVs is considered by each UAV.

2) The proposed game is proven to be an exact potential game. The existence of the Stackelberg equilibrium (SE) is also proved according to the properties of the exact potential game and the limited strategic game.

3) A distributed cooperative anti-jamming channel selection algorithm based on local altruistic is proposed to obtain the final SE. Simulation results show that the algorithm has a fast convergence rate and is close to the optimal Nash equilibrium (NE). The proposed cooperative anti-jamming method achieves better performance compared with non-cooperative ones.

The rest of this paper is organized as follows. The system model and game model is described in Section 2. The proposed algorithm of cooperative anti-jamming is developed in Section 3. In Section 4, the simulation results are shown and analyzed. Finally, some conclusions are drawn in Section 5.

Related notations are shown in Table 1.

Table 1 Notations.

| notations | meaning |
| --- | --- |
| $h_n$ | flight altitude of UAV $n$ |
| $(x_{n,z}, y_{n,z})$ | horizontal coordinate of UAV $n$ in period $z$ |
| $(x_j, y_j)$ | horizontal coordinate of the jammer |
| $(x_0, y_0)$ | horizontal coordinate of the FC |
| $\mathcal{N}$ | UAVs set |
| $\mathcal{M}$ | available channels set |
| $\mathbf{a}_z$ | channel selection strategy of UAVs |
| $c_{j,z}$ | channel selection of the jammer |
| $a_{n,z}$ | channel selected by UAV $n$ in period $z$ |
| $c_{j,z}$ | channel selected by the jammer in period $z$ |
| $H_{n,z}^{(a_{n,z})}$ | instantaneous channel gain between UAV $n$ and the FC |
| $H_j^{(c_{j,z})}$ | instantaneous channel gain between the jammer and the FC |
| $p_n$ | transmission power of UAV $n$ |
| $p_j$ | transmission power of the jammer |

## 2. System model and problem formulation

We consider that a cluster of UAVs are executing telemetry tasks and sending information to the FC, as is shown in Fig. 1. To prevent collisions, the flight altitude of UAV $n$ is set as $h_n$. The trajectory from the beginning to the destination is divided into $Z$ periods, and each period has time $T$. The coordinate of period $z \in \{1,2,...,Z\}$ is defined as $(x_{n,z}, y_{n,z}, H_n)$. The intelligent jammer located on $(x_j, y_j)$ sends jamming signals to the FC on $(x_0, y_0)$. The location of the jammer can be estimated with a high accuracy by the localization techniques.[16] It can sense the channel available to UAVs, and adaptively adjust its strategy to achieve the maximum jamming effect. On the other hand, UAVs are intelligent and choose channels flexibly to minimize external interference and co-channel interference received.

The UAVs set is denoted as $\mathcal{N} = \{1,2,\cdots,N\}$, and the available channels set is denoted as $\mathcal{M} = \{1,2,\cdots,M\}$. The channel selection strategy of all UAVs in period $z$ is set as $\mathbf{a}_z = \{a_{1,z}, a_{2,z}, \cdots, a_{N,z}\}$, $a_{n,z} \in \mathcal{M}$, The channel selection strategy of all UAVs except UAV $n$ is expressed as $\mathbf{a}_{-n,z} = \{a_{1,z}, \cdots, a_{n-1,z}, a_{n+1,z}, \cdots, a_{N,z}\}$. The available channels of jammer is set as $\mathcal{C} = \{c_1, c_2, \cdots, c_J\}$, and the jamming channel is $c_{j,z} \in \mathcal{C}$.

The information transmission process of UAVs is shown in Fig. 2. The UAVs and FC can obtain the perfect channel state information (CSI) through channel estimation during the channel coherence time.[17] Therefore, it is supposed that the CSI is known. Channel selection can be made quickly, and the complexity is given in Section 3. The channel estimation and selection are completed in a short period of time defined as $\tau$. After that, data is transmitted in $T - \tau$.

Due to the moving transmitter and varying channel characteristics of UAVs, the UAV-to-FC channel is different from traditional channels.[18,19] Assume that the wireless channel experiences block fading[20], that is, in one period, the characteristics of the channel remain constant and change randomly in the next period. The channel model between UAV $n$ and the FC is given by $H_{n,z}^{(a_{n,z})}$.

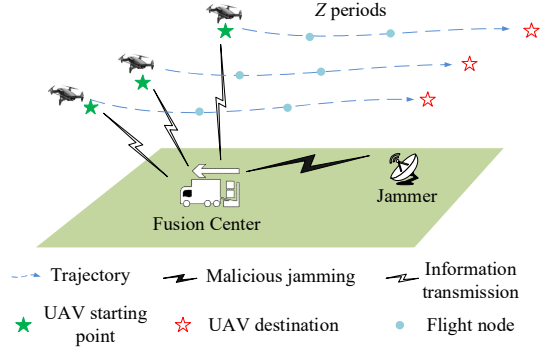

Fig. 1.  System model.

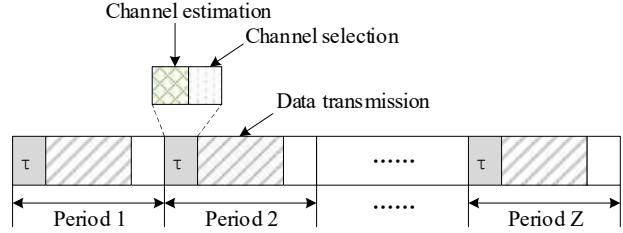

Fig. 2.  Process of information transmission in time domain.

It is considered that the channel between the jammer and the FC follows the Rayleigh fading model. Because the jammer and the FC cannot conduct channel training, the channel models established by the two sides are different. Without loss of generality, the jammer-to-FC small-scale fading gain modeled by the jammer and the FC are set as $\mathbf{g}_a = \{g_{a,1}, g_{a,2}, \cdots, g_{a,s}, \cdots, g_{a,S}\}$ and $\mathbf{g}_j = \{g_{j,1}, g_{j,2}, \cdots, g_{j,s}, \cdots, g_{j,S}\}$, respectively, where $s \in \{1, 2, \cdots, S\}$ is the sample index, $g_{j,s}$ and $g_{a,s}$ are the small-scale fading gains following the probabilities $\mathbf{Pr}_a = \{\Pr_{a,1}, \Pr_{a,2}, \cdots, \Pr_{a,s}, \cdots, \Pr_{a,S}\}$, $\sum_{s=1}^{S} \Pr_{a,s} = 1$ and $\mathbf{Pr}_j = \{\Pr_{j,1}, \Pr_{j,2}, \cdots, \Pr_{j,s}, \cdots, \Pr_{j,S}\}$, $\sum_{s=1}^{S} \Pr_{j,s} = 1$. The expected jammer-to-FC channel gains modeled by the jammer and the FC are given as:

$$H_a^{(c_{j,z})} = \overline{g}_{a,s} \left( d_j \right)^{-\alpha} \tag{1}$$

$$H_j^{(c_{j,z})} = \overline{g}_{j,s} \left( d_j \right)^{-\alpha} \tag{2}$$

where $d_j = \sqrt{(x_0 - x_j)^2 + (y_0 - y_j)^2}$ is the distance between the jammer and the FC, $\alpha$ is the path loss coefficient; and $\overline{g}_{j,s} = \mathbf{g}_j \cdot \mathbf{Pr}_j^{\mathrm{T}}$, $\overline{g}_{a,s} = \mathbf{g}_a \cdot \mathbf{Pr}_a^{\mathrm{T}}$ are the expected channel gains.

In period $z$, the signal to interference plus noise ratio (SINR) of UAV $n$ to the FC is as follows,

$$\gamma_{n,z} = \frac{p_n H_{n,z}^{(a_{n,z})}}{N_0 + I_{n,z} + J_{n,z}} \tag{3}$$

where $p_n$ is the transmission power of the UAV $n$; $N_0$ is the noise power spectral density; the sum mutual interference caused by all other UAVs is denoted as $I_{n,z} = \sum_{k \in \{\mathcal{N} / n\}} p_k H_{k,z}^{(a_{k,z})} f(a_{n,z}, a_{k,z})$; and the malicious interference is $J_{n,z} = p_j H_j^{(c_{j,z})} f(a_{n,z}, c_{j,z})$ with the jamming power $p_j$.

To improve the network performance, effective anti-jamming channel selection is faced with the following challenges. That is, it is essential to reduce malicious jamming as well as mitigate co-channel interference among UAVs. Inspired by Ref. 21, we consider that less interference of the UAV is equivalent to higher throughput.

Due to the limited UAVs' payload, the battery capacity is finite. [22] Therefore, we consider minimizing the loss value under a certain flight energy cost. The loss of the UAV $n$ in period $z$ is described as expected weighted sum interference (i.e., the first term and the second term in Eq. 4) plus flight energy consumption (i.e., the third term in

Eq. 4) which is non-negligible for UAVs, as follows,

$$E_{n,z} = p_n p_j H_j^{(c_{j,z})} f(a_{n,z}, c_{j,z}) + \sum_{k \in \{\mathcal{N}/n\}} p_n p_k H_{k,z}^{(a_{k,z})} f(a_{n,z}, a_{k,z}) + C_f C_0 d_{n,z,z-1} \qquad (4)$$

where $C_f$ is the flight energy consumption per unit distance; $C_0$ is a constant factor to balance the weighted interference; jamming with the flight cost, and $d_{n,z,z-1}$ is the flight distance of UAV $n$ in period $z$; and $f(\cdot)$ is an indicator function as

$$f(x,y) = \begin{cases} 1, & x = y \\ 0, & x \neq y \end{cases} \qquad (5)$$

Based on the above analysis, the objective of UAVs is to optimize the channel selection strategies in each period to minimize the total system loss. Therefore, the optimization problem can be described as

$$P : \mathbf{a}_{opt} = \arg \min \sum_{n}^{N} E_{n,z} \qquad (6)$$

$$s.t. \quad n = 1, 2, ..., N. \ z = 1, 2, ..., Z$$

where $\mathbf{a}_{opt}$ is the optimal combination of strategies for all UAVs. This problem is a combinatorial optimization problem, the computational complexity of centralized exhaustive method is too high, and the convergence of heuristic method is difficult to ensure. Therefore, we model the scene as a Stackelberg game, and design an effective distributed method based on local altruistic to obtain the game equilibrium.

### 2.1. Stackelberg game formulation

Inspired by Ref. 23, a cooperative anti-jamming channel selection method based on a local altruistic game is proposed to solve the above problem. The game model is represented as $\mathcal{G} = \{\mathcal{N}, \mathcal{J}, \mathcal{M}, \mathcal{C}, u_{n,z}, u_{j,z}\}$ where $\mathcal{N}$ and $\mathcal{J}$ are the UAV set and jammer, respectively. $\mathcal{M}$ and $\mathcal{C}$ are the strategy space of UAVs and the jammer, respectively, $u_{n,z}$ and $u_{j,z}$ are the corresponding utilities. Denote UAVs as followers and the jammer as the leader in the Stackelberg game.

Consider local altruistic behavior among UAVs, in which each UAV considers the utility of both itself and all other UAVs. The utility function of UAV $n$ in period $z$ is defined as

$$u_{n,z}(a_{n,z}, \mathbf{a}_{-n,z}, c_{j,z}) = W - (E_{n,z} + \sum_{k \in \{\mathcal{N}/n\}} E_{k,z}) \qquad (7)$$

where $W$ is a predefined positive constant to guarantee that the utility is not negative, $E_{n,z}$ is the loss value of UAV $n$, $\sum_{k \in \{\mathcal{N}/n\}} E_{k,z}$ is the sum loss value of all other UAVs. Therefore, the optimization problem of UAV $n$ is as

$$a_{n,z} = \arg \max_{a_{n,z}} u_{n,z}(a_{n,z}, \mathbf{a}_{-n,z}, c_{j,z}) \qquad (8)$$

The subgame of the follower is defined as

$$\mathcal{G}_f = \{\mathcal{N}, \mathcal{M}, u_{n,z}\} \qquad (9)$$

where $\mathcal{N}$ is the UAV set, and $\mathcal{M}$ is the strategy set of UAVs. Similarly, the utility function and the optimization problem of the jammer are respectively

$$u_{j,z} = \sum_{n=1}^{N} p_n p_j H_a^{(c_{j,z})} f(a_{n,z}, c_{j,z}) \qquad (10)$$

$$c_{j,z} = \arg \max u_{j,z}(\mathbf{a}_z, c_{j,z}) \qquad (11)$$

The subgame of the leader is defined as

$$\mathcal{G}_l = \left\{ \mathcal{J}, \mathcal{C}, u_{j,z} \right\} \tag{12}$$

where the $\mathcal{J}$ is the jammer, $\mathcal{C}$ is jammer's strategy set with a cardinality of $\mathcal{M}$.

### 2.2. Analysis of Stackelberg equilibrium

**Definition 1 (Stackelberg Equilibrium):** The Stackelberg Equilibrium $\left( \mathbf{a}_z^*, c_{j,z}^* \right)$ is defined as follows[24],

$$u_j \left( \mathbf{a}_z^*, c_{j,z}^* \right) \geq u_j \left( \mathbf{a}_z^*, c_{j,z} \right) \tag{14}$$

$$u_n \left( a_{n,z}^*, \mathbf{a}_{-n,z}^*, c_{j,z}^* \right) \geq_n \left( a_{n,z}, \mathbf{a}_{-n,z}^*, c_{j,z}^* \right) \tag{15}$$

**Definition 2 (Exact Potential Game):** A game $\mathcal{G}$ is an exact potential game if there exists a potential function $\Phi$ such that[25]

$$\Phi \left( \tilde{a}_{n,z}, \mathbf{a}_{-n,z}, c_{j,z} \right) - \Phi \left( a_{n,z}, \mathbf{a}_{-n,z}, c_{j,z} \right) = u_n \left( \tilde{a}_{n,z}, \mathbf{a}_{-n,z}, c_{j,z} \right) - u_n \left( a_{n,z}, \mathbf{a}_{-n,z}, c_{j,z} \right) \tag{13}$$

where $\tilde{a}_n$ the changed action of UAV $n$. The exact potential game has two most important properties, as follows:

1) There is at least a pure strategy NE;
2) The global or local optimal solution of the potential function is a NE.

**Theorem 1:** For a given jammer's strategy, the subgame of the follower is an exact potential game, and there is at least a pure strategy NE. The optimal solution to the problem of minimizing the total loss of the system is the pure strategy NE of the game.

*Proof:* See Appendix.

**Theorem 2:** There exists a jammer's stationary policy and a user's NE that constitute a SE in the proposed game.

*Proof:* Based on Theorem 1, the subgame of the follower is an exact potential game, and there is at least a pure strategy NE.

For the subgame of the leader, it is a finite strategy game. According to Ref. 26, if a game is a finite strategy game, there must be a mixed strategy equilibrium. Therefore, there exists a stationary SE in the proposed game.

Inspired by Ref. 27, given the jammer's strategy, there always exists $\mathrm{NE}\left( c_{j,z} \right)$ in the proposed game. The jammer's stationary strategy can be expressed as

$$c_{j,z}^* = \arg \max_{c_{j,z}} u_j \left( c_{j,z}, NE \left( c_{j,z} \right) \right) \tag{16}$$

According to Ref. 26, every finite strategy game has a mixed strategy equilibrium. Therefore $\left( c_{j,z}^*, NE \left( c_{j,z}^* \right) \right)$ constitutes a stationary equilibrium. ∎

The game model proposed in this paper is an exact potential game and has good convergence behaviors. In other words, since the utility function changes by the same amount as the potential function, when one player adjusts its strategy in the direction of increasing the utility function, this adjustment also increases the potential function and eventually converge to the pure strategy NE of the game in a finite number of iterations.

## 3. Distributed algorithm

### 3.1. Algorithm description

To solve the proposed game, the cooperative anti-jamming channel selection game is extended to a mixed strategy form. UAV $n$'s mixed strategy in slot $t$ is defined as $\boldsymbol{\theta}_n(t) = \left( \theta_{n1}(t), \cdots, \theta_{nm}(t), \cdots, \theta_{nM}(t) \right)$, $\sum_{m \in \mathcal{M}} \theta_{nm}(t) = 1$. The probability that UAV $n$ chooses channel $m$ in slot $t$ is set as $\theta_{nm}(t)$. Similarly, the jammer's mixed strategy in epoch $k$ is defined as $\boldsymbol{\theta}_j(k) = \left( \theta_{j1}(k), \cdots, \theta_{jm}(k), \cdots, \theta_{jM}(k) \right)$, $\sum_{m \in \mathcal{M}} \theta_{jm}(k) = 1$. The probability that the jammer chooses channel $m$ in epoch $k$ is set as $\theta_{0m}(k)$.

Suppose the channel selection strategy of all UAVs in slot $t$ is $\mathbf{a}(t) = \{a_1(t), a_2(t), \cdots, a_N(t)\}$ and the jammer's channel is $c_j(t)$. The utility function of UAV $n$ in slot $t$ can be formulated as follows:

$$
\begin{aligned}
u_n(t) = W - &\left( E_{n,z} + \sum_{i \in \{\mathcal{N}/n\}} E_{i,z} \right) \\
= W - &\left( \sum_{m \in \{\mathcal{N}/n\}} p_n p_m H_{m,z}^{(a_{m,z})} f\left(a_n(t), a_m(t)\right) + p_n p_j H_j^{(c_{j,z})} f\left(a_n(t), c_j(t)\right) + C_f C_0 d_{n,z,z-1} \right. \\
& \left. + \sum_{i \in \{\mathcal{N}/n\}} \left( \sum_{m \in \{\mathcal{N}/i\}} p_i p_m H_{m,z}^{(a_{m,z})} f\left(a_i(t), a_m(t)\right) + p_i p_j H_j^{(c_{j,z})} f\left(a_i(t), c_j(t)\right) + C_f C_0 d_{i,z,z-1} \right) \right)
\end{aligned}
\tag{17}
$$

where $W$ is a positive constant to guarantee $u_n(t)$ is not negative.

The utility function of the jammer in epoch $k$ is as

$$
u_j(k) = \sum_{n \in \mathcal{N}} p_n p_j H_a^{(c_{j,z})} f\left(c_j(k), a_n(k)\right)
\tag{18}
$$

Details of cooperative anti-jamming algorithm based on SLA are given in **Algorithm 1**. The termination condition of the algorithm is to reach the maximum number of iterations or the channel selection probability satisfies the condition $\theta_{jm}(k) > Q$, $\forall m \in M$ where Q is a convergence threshold.

---

***Algorithm 1:* Cooperative Anti-Jamming Algorithm Based on SLA**

---

**Initialization:** Set $t = 0$, $k = 0$. The initial channel selection probability is set as $\theta_{0m}(k) = \theta_{nm}(t) = \dfrac{1}{|\mathcal{M}|}$, $\forall m \in \mathcal{M}$.

**Step 1:** In the epoch $k$, the jammer chooses a channel $c_j(k)$ to interfere according to $\boldsymbol{\theta}_0(k)$.

**Step 2:** In the epoch $k$, the learning process of each UAV's strategy is as follows:

(1) In the slot $t$, UAV $n$ chooses a channel $a_n(t)$ according to $\boldsymbol{\theta}_n(t)$.

(2) UAV $n$ calculates its utility $u_n(t)$ in (17).

(3) Each user updates the channel selection probability according to the following rules:

$$\theta_{nm}(t+1) = \theta_{nm}(t) + b_1 \tilde{u}_n(t)\left(1 - \theta_{nm}(t)\right), m = a_n(t)$$
$$\theta_{nm}(t+1) = \theta_{nm}(t) - b_1 \tilde{u}_n(t)\theta_{nm}(t), \qquad m \neq a_n(t)$$

where $0 < b_1 < 1$ is the learning step size of UAVs, $\tilde{u}_n(t)$ is the normalized utility.

**Step 3:** The jammer chooses a channel $c_j(k)$ according to $\boldsymbol{\theta}_0(k)$;

**Step 4:** The jammer calculates its utility $u_j(k)$ in (18).

**Step 5:** The jammer updates the channel selection probability according to the following rules:

$$\theta_{jm}(k+1) = \theta_{jm}(k) + b_2 \tilde{u}_j(k)\left(1 - \theta_{jm}(k)\right), \ m = c_j(k)$$
$$\theta_{jm}(k+1) = \theta_{jm}(k) - b_2 \tilde{u}_j(k)\theta_{jm}(k), \qquad m \neq c_j(k)$$

where $0 < b_2 < 1$ is the learning step size of the jammer, $\tilde{u}_j(k)$ is the normalized utility.

**Step 6:** $k = k + 1$, and go to **Step 1** until the termination condition is met.

---

*3.2. Analysis of algorithm*

**Theorem 3:** With sufficiently small step sizes $b_1$ and $b_2$, **Algorithm 1** converges to a NE in the follower subgame.

Proof: According to **Theorem 1**, the follower subgame is an exact potential game. Then by following Theorem 5 in Ref. 15, the SLA based learning algorithm can be proven to converge towards NE. ∎

It is remarkable that the channel selection probability is updated on the basis of the payoff feedback. The probability of choosing the channel increases if the payoff is received positively. In addition, the smaller the value of step sizes $b_1$ and $b_2$ are, the more accurate it is. But meanwhile the calculation speed decrease. We can get the appropriate values through simulation experiments.

The analysis of the computational complexity of the proposed algorithm is provided in Table 2.

1) Denote the channel training and CSI broadcasting process by $C_1$, which is a small constant. The complexity is $\mathcal{O}(C_1)$.

2) Secondly, UAVs sense their aggregated weighted interference simultaneously in a small sensing duration $C_2$. The complexity of this period can be donated as $\mathcal{O}(C_2)$.

3) Then the complexity of computing UAVs' respective payoffs can be expressed as $\mathcal{O}(C_3)$, where $C_3$ is also a short duration.

4) After computing payoffs, UAV $n$ update its channel selection probability immediately, and the complexity is $\mathcal{O}(C_4)$.

5) Finally, UAV $n$ select the channel according to the converged probability for one time slot.

Table 2 Complexity of collaborative anti-jamming algorithm based on SLA.

| Operations | Complexity |
|---|---|
| estimate channels and broadcast CSI | $\mathcal{O}(C_1)$ |
| sense aggregated interference | $\mathcal{O}(C_2)$ |
| calculate payoffs | $\mathcal{O}(C_3)$ |
| update channel selection probability | $\mathcal{O}(C_4)$ |
| assign UAVs to select channels | 1 |
| total complexity | $\mathcal{O}(C_1)+\mathcal{O}(C_2)+\mathcal{O}(C_3)+\mathcal{O}(C_4)+1$ |

To sum up, the total complexity of SLA-based learning algorithm is about $\mathcal{O}(C_1)+\mathcal{O}(C_2)+\mathcal{O}(C_3)+\mathcal{O}(C_4)+1$.

## 4. Simulation results and discussions

To verify the proposed method, simulation results are given and discussed in this section. Some basic settings are as follows. As depicted in Fig. 3, the mission area is 200m×100m with six UAVs and one jammer distributed in it. UAVs keep flying at different altitudes which constitute a vector $\mathbf{H} = [100, 110, 120, 130, 140, 150]$ m. Let UAVs take off from their respective starting points to destinations. The overall flight duration is discretized into 6 periods of the same length. The FC and the jammer are located at [100, 140] m and [120, 70] m on the ground, respectively. The parameters are set as Table 3.

Table 3 Parameters.

| Parameters | Value | Parameters | Value |
|---|---|---|---|
| $d_0$ | 50 m | $c_f$ | 1 |
| $N_0$ | -70 dB | $C_0$ | $10^{-3}$ |
| $p_n$ | 10 W | $\mathbf{g}_a$ | {0.5, 0.8, 1, 1.5, 2} |
| $p_j$ | 30 W | $\mathbf{g}_j$ | {0.5, 1, 1.5, 2, 2.5} |

| $\alpha$ | 2 | $\mathbf{Pr}_a$ | {0.21, 0.22, 0.14, 0.28, 0.15} |
| $b_1$ | 0.2 | $\mathbf{Pr}_j$ | {0.14, 0.28, 0.28, 0.18, 0.12} |
| $b_2$ | 0.3 | $W$ | 1 |
| $H_{n,z}^{(a_n,z)}$ | $1.1 \times d_{n,z}^{-2}$ | | |

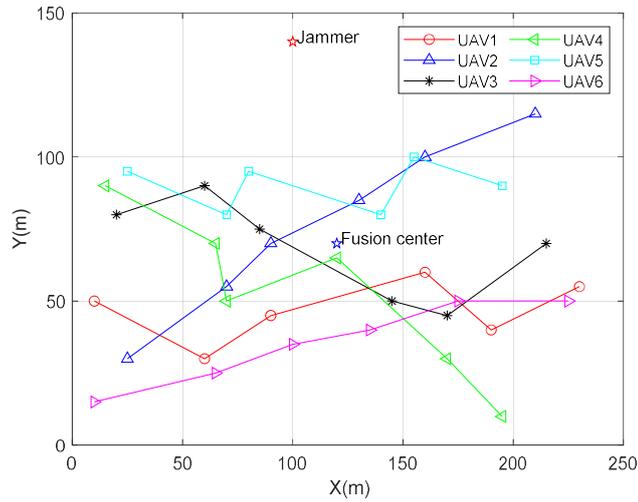

Fig. 3.   Scenarios.

### 4.1. Convergence performance

The resulting channel selection policies of the UAVs and the jammer in each period are shown in Fig. 4. It can be seen that UAVs who are close to each other, such as UAV 1 and 6, select different channels in periods 1-6, while UAVs who are far away, such as UAV 1 and 5, share same channels in periods 1-3. This is because the closer the UAVs are, the stronger the interference is; otherwise, the weaker the interference is. On the other hand, since the jammer makes the decision after the user, to maximize the jamming effect, the jammer chooses the channel which is accessed by the most UAVs that are close to the jammer. This verifies the intelligence of the jammer.

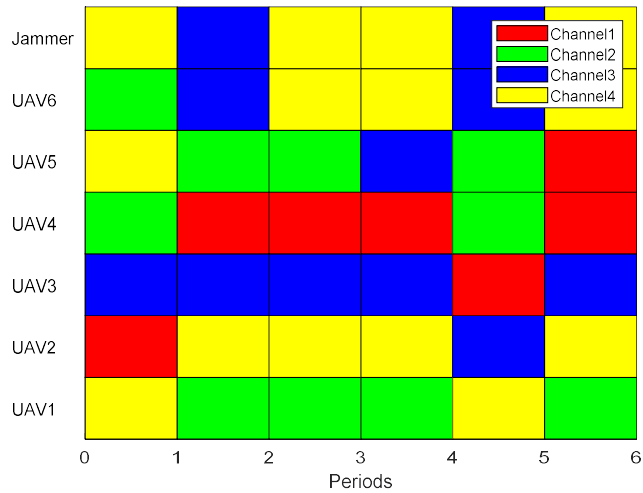

Fig. 4.   Channel selection of UAVs and the jammer.

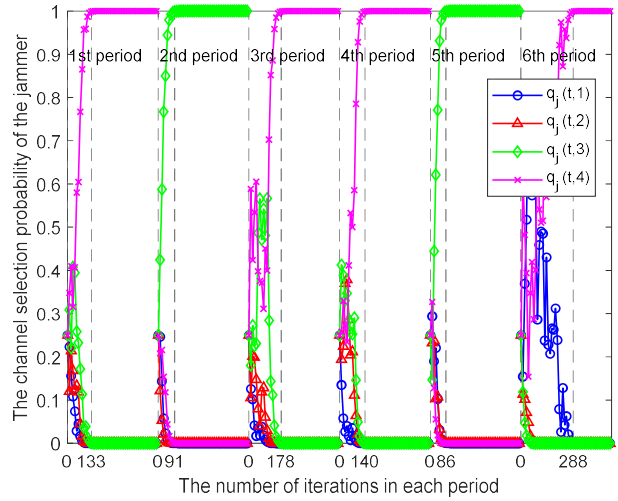

Fig. 5.    Convergence of channel selection probability of the jammer.

The convergence of channels choice probability of the jammer in different periods is shown in Fig. 5, where $q_j(z, c)$ is the probability of the jammer choosing channel $c$ in period $z$. Iteration is stopped if a preset objective function tolerance (i.e., the probability of choosing the channel > 0.999) or the maximum number of iterations 500 is met. The jammer selects channels 4, 3, 4, 4, 3, and 4 in turn in 6 periods, which is consistent with the channel selection of the jammer in Fig. 4. The jammer's channel selection probabilities can converge in less than 290 epochs, which indicates that the proposed algorithm has a fast convergence rate.

To analyze the convergence of UAVs' channel selection, take UAV 2 as an example. Fig. 6 shows the convergence curve of UAV 2's channel selection probability in different periods, where $q_2(z, c)$ is the probability of UAV 2 choosing channel $c$ in period $z$. The maximum number of iterations is 300. UAV 2 selects channels 1, 4, 4, 4, 3, and 4 in turn in 6 periods, which is consistent with the channel selection of UAV 2 in Fig. 4. The UAV 2's channel selection probabilities can converge in less than 100 slots, which indicates that the proposed algorithm has a fast convergence rate.

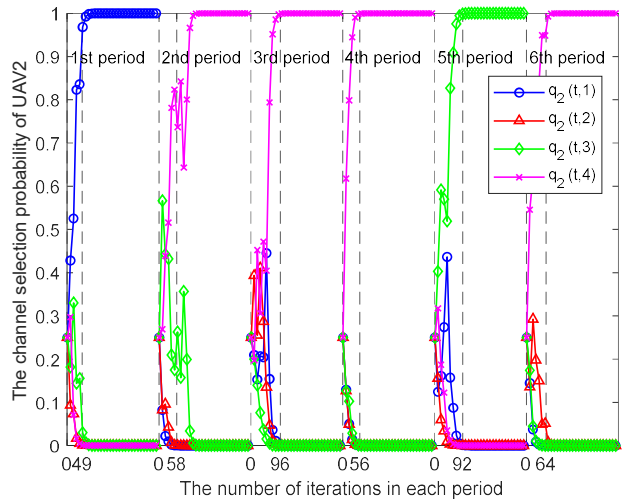

Fig. 6.    Convergence of channel selection probability of UAV 2.

### 4.2. Performance Comparison

To evaluate the performance of the proposed algorithm, the comparison curves of total system loss with the proposed algorithm, the best NE, the worst NE and the random selection are presented in Fig. 7. As the number of channels increases, the total system loss of these algorithms all decreases. This is because the mutual interference between users decreases. Compared with the random selection algorithm, the total network loss of the proposed algorithm decreases obviously, especially when the number of channels is 6, the anti-jamming performance increases by about 64%. Additionally, when the number of channels is 2, the anti-jamming performance of our proposed algorithm is about 32% better than that of non-cooperative ones, which shows the superiority of the cooperative algorithm. The proposed algorithm is close to the best NE, which indicates the reliability of the proposed algorithm.

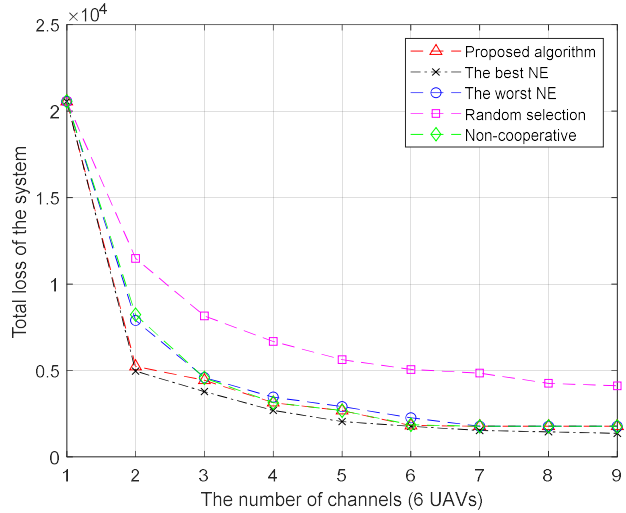

Fig. 7.   The loss of the total system of different algorithms.

Fig. 8 shows the impact of transmission power of the jammer and UAVs on the total system loss. As the power of the jammer increases, the interference becomes more serious, leading to the increase of the total loss of the system. At the same time, the increase of users' power also leads to the aggravation of UAVs' interference and the increase of total interference.

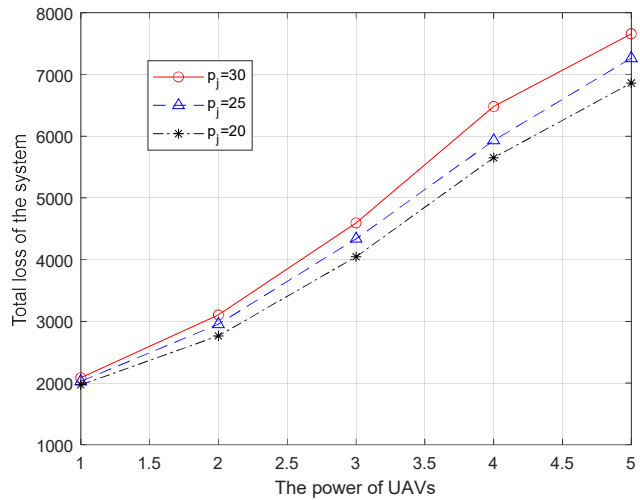

Fig. 8.   The total loss of the system versus transmission power of the jammer and UAVs.

## 5.  Conclusion

In this paper, a cooperative anti-jamming channel selection problem was studied in the multi-UAV communication scenario. The impact of malicious jamming and co-channel interference on system performance was considered. The problem was modeled as a Stackelberg game, and the local altruistic game was established among UAVs. In our game model, the UAV pursued the maximization of the utilities of itself and all other UAVs, and the jammer pursued the maximization of its own utility. To obtain the game equilibrium, a distributed cooperative anti-jamming algorithm based on local altruism was proposed. Finally, the simulation results showed that the proposed algorithm satisfied good convergence, and significantly improved the anti-jamming performance.

## References

1. Zeng Y, Zhang R, Lim TJ. Wireless Communications with Unmanned Aerial Vehicles: Opportunities and Chal-

lenges. *IEEE Communications Magazine* 2016;54(5):36-42.

2. Kumar N, Puthal D, Theocharides T, Mohanty SP. Unmanned Aerial Vehicles in Consumer Applications: New Applications in Current and Future Smart Environments. *IEEE Consumer Electronics Magazine* 2019;8(3):66-7.

3. Sun W, Liu J, Zhang H. When smart wearables meet intelligent vehicles: challenges and future directions, *IEEE Wireless Commun* 2017;24(3):58-65.

4. Xu Y, Ren G, Chen J, et al. A one-leader multi-follower Bayesian-Stackelberg game for anti-jamming transmission in UAV communication networks. *IEEE Access* 2018;6:21697-709.

5. Zheng J, Cai Y, Lu N, et al. Stochastic game-theoretic spectrum access in distributed and dynamic environment. *IEEE Transactions on Vehicular Technology* 2015;64(10):4807-20.

6. Duan B, Yin D, Cong Y, et al. Anti-jamming path planning for unmanned aerial vehicles with imperfect jammer information. *Proceedings of 2018 IEEE International Conference on Robotics and Biomimetics (ROBIO)*; 2018 Dec 12-15; Kuala Lumpur, Malaysia. IEEE; 2018. p. 729-35.

7. Wang HC, Wang JL, Ding GR, Chen J. Intelligent cooperative anti-jamming technology in space air-ground integrated networks. *Journal of Command and Control* 2020;6(3):185-91.

8. Dai C, Xu D, Xiao L, et al. Collaborative UFH-based anti-jamming broadcast with learning. *Proceedings of 2017 IEEE/CIC International Conference on Communications in China (ICCC)*; 2017 Oct 22-24; Qingdao, China. IEEE press; 2017. p. 1-5.

9. Chen G, Li Y, Xiao L, et al. Collaborative anti-jamming broadcast with uncoordinated frequency hopping over USRP. *Proceedings of 2015 IEEE 81st Vehicular Technology Conference (VTC Spring)*; 2015 May11-14; Glasgow, UK. IEEE press; 2015. p. 1–6.

10. Ahmed I K, Fapojuwo A O. Stackelberg equilibria of an anti-jamming game in cooperative cognitive radio networks. *IEEE Transactions on Cognitive Communications and Networking* 2017;4(1):121-34.

11. Zhang Y, Xu Y, Xu Y, et al. A multi-leader one-follower Stackelberg game approach for cooperative anti-jamming: no pains, no gains. *IEEE Communications Letters* 2018;22(8):1680-3.

12. Yao F, Jia L. A collaborative multi-agent reinforcement learning anti-jamming algorithm in wireless networks. *IEEE Wireless Communications Letters* 2019;8(4):1024-7.

13. Xu Y, Chen J, Xu Y, et al; inventors; Beijing power patent agency, assignee. Multi-user cooperative anti-jamming system and dynamic spectrum cooperative anti-jamming method. China patent CN 112188504A. 2021 Jan 5 [Chinese].

14. Wang H, Li J, Zhao H, et al. Anti-jamming network architecture self-adaption technology based on cooperation and cognition. *Journal of Computer Applications* 2016;36(9):2367-73 [Chinese].

15. Xu Y, Wang J, Wu Q, et al. Opportunistic spectrum access in unknown dynamic environment: A game-theoretic stochastic learning solution. *IEEE Transactions on Wireless Communications* 2012;11(4):1380-91.

16. Li J, He Y, Zhang X, et al. Simultaneous localization of multiple unknown emitters based on UAV monitoring big data. *IEEE Transactions on Industrial Informatics* 2021 17(9):6303-13.

17. Bogale T. E., Le L. B, Wang X. Hybrid analog-digital channel estimation and beamforming: training-throughput tradeoff. *IEEE Transactions on Communications* 2015;63(12):5235-49.

18. Zhu Q, Yang Y, Wang C, et al. Spatial correlations of a 3-D non-stationary MIMO channel model with 3-D antenna arrays and 3-D arbitrary trajectories. *IEEE Wireless Communications Letters* 2019;8(2):512-5.

19. Zhu Q, Li H, Fu Y, et al. A Novel 3D Non-stationary Wireless MIMO Channel Simulator and Hardware Emulator. *IEEE Transactions on Communications* 2018;66(9):3865-78.

20. Wu Q, Xu Y, Wang J, et al. Distributed channel selection in time-varying radio environment: interference mitigation game with uncoupled stochastic learning. *IEEE Transactions on Vehicular Technology* 2013;62(9):4524-38.

21. Zheng J, Cai Y, Lu N, et al. Stochastic game-theoretic spectrum access in distributed and dynamic environment. *IEEE Transactions on Vehicular Technology* 2015;64(10):4807-20.

22. Qi N, Wang M, Wang W, et al. Energy efficient full-duplex UAV relaying networks under load-carry-and-delivery scheme. *IEEE Access* 2020;8:74349-58.

23. Xu Y, Wang J, Wu Q, Anpalagan A, et al. Opportunistic spectrum access in cognitive radio networks: Global optimization using local interaction games. *IEEE Journal of Selected Topics in Signal Processing* 2012;6(2):180-94.

24. L. Jia, F. Yao, Y. Sun, et al. Bayesian Stackelberg game for antijamming transmission with incomplete information. *IEEE Communications Letters* 2016;20(10):1991-4.

25. Monderer D, Shapley LS. Potential games. *Games and Economic Behavior* 1996;14(1):124-43.

26. Han Z, Niyato D, Saad W, et al. *Game theory in wireless and communication networks*. Cambridge University Press, 2012.

27. Chen X, Zhang H, Chen T, et al. Improving energy efficiency in green femtocell networks: A hierarchical reinforcement learning framework. *Proceedings of 2013 IEEE International Conference on Communications (ICC)*; 2013 Jun 9-13; Budapest, Hungary. IEEE press; 2013. p. 2241-5.

## Appendix

Here, we provide the proof for the Theorem 1. Given the jammer's strategy, the potential function of the subgame of the follower can be constructed as

$$
\begin{aligned}
\Phi\left(a_{n,z}, \mathbf{a}_{-n,z}, c_{j,z}\right) &= \Phi_1\left(a_{n,z}, \mathbf{a}_{-n,z}, c_{j,z}\right) + \Phi_2\left(a_{n,z}, \mathbf{a}_{-n,z}, c_{j,z}\right) + \Phi_3\left(d_{n,z,z-1}\right) \\
&= \sum_{n=1}^{N}\sum_{k\in\{\mathcal{N}/n\}} p_n p_k H_{k,z}^{(a_{k,z})} f\left(a_{n,z}, a_{k,z}\right) + \sum_{n=1}^{N} p_n p_j H_j^{(c_{j,z})} f\left(a_{n,z}, c_{j,z}\right) + C_f C_0 d_{n,z,z-1}
\end{aligned}
\tag{19}
$$

where $\Phi_1\left(a_{n,z}, \mathbf{a}_{-n,z}, c_{j,z}\right)$ can be expressed as follows,

$$
\begin{aligned}
\Phi_1\left(a_{n,z}, \mathbf{a}_{-n,z}, c_{j,z}\right) &= \sum_{n=1}^{N}\sum_{k\in\{\mathcal{N}/n\}} p_n p_k H_{k,z}^{(a_{k,z})} f\left(a_{n,z}, a_{k,z}\right) \\
&= \sum_{k\in\{\mathcal{N}/n\}} p_n p_k H_{k,z}^{(a_{k,z})} f\left(a_{n,z}, a_{k,z}\right) + \sum_{i\in\{\mathcal{N}/n\}}\sum_{k\in\{\mathcal{N}/i\}} p_i p_k H_{k,z}^{(a_{k,z})} f\left(a_{i,z}, a_{k,z}\right)
\end{aligned}
\tag{20}
$$

Let $A = \sum_{i\in\{\mathcal{N}/n\}}\sum_{k\in\{\mathcal{N}/i\}} p_i p_k H_{k,z}^{(a_{k,z})} f\left(a_{i,z}, a_{k,z}\right)$, we have

$$
A = \sum_{i\in\{\mathcal{N}/n\}}\sum_{k\in\{\mathcal{N}/\{n,i\}\}} p_i p_k H_{k,z}^{(a_{k,z})} f\left(a_{i,z}, a_{k,z}\right) + \sum_{n\in\{\mathcal{N}/k\}} p_i p_n H_{n,z}^{(a_{n,z})} f\left(a_{i,z}, a_{n,z}\right)
\tag{21}
$$

And $\Phi_2\left(a_{n,z}, \mathbf{a}_{-n,z}, c_{j,z}\right)$ can be expressed as

$$
\begin{aligned}
\Phi_2\left(a_{n,z}, \mathbf{a}_{-n,z}, c_{j,z}\right) &= \sum_{n=1}^{N} p_n p_j H_j^{(c_{j,z})} f(a_{n,z}, c_{j,z}) \\
&= p_n p_j H_j^{(c_{j,z})} f(a_{n,z}, c_{j,z}) + \sum_{i\in\{\mathcal{N}/n\}} p_i p_j H_j^{(c_{j,z})} f(a_{i,z}, c_{j,z})
\end{aligned}
\tag{22}
$$

where $\sum_{i\in\{\mathcal{N}/n\}} p_i p_j H_j^{(c_{j,z})} f(a_{i,z}, c_{j,z})$ is irrelevant to $a_{n,z}$.

Notice that $\Phi_3(d_{n,z,z-1}) = C_f C_0 d_{n,z,z-1}$, a function about distance, is irrelevant to $a_{n,z}$. Thus, when UAV $n$ unilaterally changes action $a_{n,z}$ to $\tilde{a}_{n,z}$, the change in utility function is as

$$
\begin{aligned}
\Delta u_{n,z} &= u_{n,z}\left(\tilde{a}_{n,z}, \mathbf{a}_{-n,z}, c_{j,z}\right) - u_{n,z}\left(a_{n,z}, \mathbf{a}_{-n,z}, c_{j,z}\right) \\
&= \left( p_n p_j H_j^{(c_{j,z})} f\left(a_{n,z}, c_{j,z}\right) + \sum_{k\in\{\mathcal{N}/n\}} p_n p_k H_{k,z}^{(a_{k,z})} f\left(a_{n,z}, a_{k,z}\right) + \sum_{k\in\{\mathcal{N}/n\}}\left( p_k p_j H_j^{(c_{j,z})} f\left(a_{k,z}, c_{j,z}\right) + \right. \right. \\
&\quad \left. \sum_{i\in\{\mathcal{N}/\{k,n\}\}} p_k p_i H_{i,z}^{(a_{i,z})} f\left(a_{k,z}, a_{i,z}\right) + p_k p_n H_{n,z}^{(a_{n,z})} f\left(a_{k,z}, a_{n,z}\right) \right) \right) - \left( p_n p_j H_j^{(c_{j,z})} f\left(\tilde{a}_{n,z}, c_{j,z}\right) + \right. \\
&\quad \sum_{k\in\{\mathcal{N}/n\}} p_n p_k H_{k,z}^{(a_{k,z})} f\left(\tilde{a}_{n,z}, a_{k,z}\right) + \sum_{k\in\{\mathcal{N}/n\}}\left( p_k p_j H_j^{(c_{j,z})} f\left(a_{k,z}, c_{j,z}\right) + \sum_{i\in\{\mathcal{N}/\{n,k\}\}} p_k p_i H_{i,z}^{(a_{i,z})} f\left(a_{k,z}, a_{i,z}\right) + \right. \\
&\quad \left.\left. p_k p_n H_{n,z}^{(\tilde{a}_{n,z})} f\left(a_{k,z}, \tilde{a}_{n,z}\right) \right)\right)
\end{aligned}
\tag{23}
$$

By combining similar terms $\sum_{k\in\{\mathcal{N}/n\}} p_k p_j H_j^{(c_{j,z})} f\left(a_{k,z}, c_{j,z}\right)$ and $\sum_{k\in\{\mathcal{N}/n\}}\sum_{i\in\{\mathcal{N}/\{n,k\}\}} p_k p_i H_{i,z}^{(a_{i,z})} f\left(a_{k,z}, a_{i,z}\right)$, Eq. 23 can be further simplified as

$$
\begin{aligned}
\Delta u_{n,z} &= \left( p_n p_j H_j^{(c_{j,z})} f\left(a_{n,z}, c_{j,z}\right) + \sum_{k\in\{\mathcal{N}/n\}} p_n p_k H_{k,z}^{(a_{k,z})} f\left(a_{n,z}, a_{k,z}\right) + \sum_{k\in\{\mathcal{N}/n\}} p_k p_n H_{n,z}^{(a_{n,z})} f\left(a_{k,z}, a_{n,z}\right) \right) \\
&\quad - \left( p_n p_j H_j^{(c_{j,z})} f\left(\tilde{a}_{n,z}, c_{j,z}\right) + \sum_{k\in\{\mathcal{N}/n\}} p_n p_k H_{k,z}^{(a_{k,z})} f\left(\tilde{a}_{n,z}, a_{k,z}\right) + \sum_{k\in\{\mathcal{N}/n\}} p_k p_n H_{n,z}^{(\tilde{a}_{n,z})} f\left(a_{k,z}, \tilde{a}_{n,z}\right) \right)
\end{aligned}
\tag{24}
$$

The change in potential function is as

$$
\begin{aligned}
\Delta\Phi &= \Phi\left(\tilde{a}_{n,z}, \mathbf{a}_{-n,z}, c_{j,z}\right) - \Phi\left(a_{n,z}, \mathbf{a}_{-n,z}, c_{j,z}\right) \\
&= \sum_{k \in \{\mathcal{N}/n\}} p_n p_k H_{k,z}^{(a_{k,z})} f\left(a_{n,z}, a_{k,z}\right) + \sum_{n \in \{\mathcal{N}/n\}} p_i p_n H_{n,z}^{(a_{n,z})} f\left(a_{i,z}, a_{n,z}\right) + p_n p_j H_{j}^{(c_{j,z})} f(a_{n,z}, c_{j,z}) \\
&\quad - \sum_{k \in \{\mathcal{N}/n\}} p_n p_k H_{k,z}^{(a_{k,z})} f\left(\tilde{a}_{n,z}, a_{k,z}\right) - \sum_{n \in \{\mathcal{N}/n\}} p_i p_n H_{n,z}^{(\tilde{a}_{n,z})} f\left(a_{i,z}, \tilde{a}_{n,z}\right) - p_n p_j H_{j}^{(c_{j,z})} f\left(\tilde{a}_{n,z}, c_{j,z}\right)
\end{aligned}
\tag{25}
$$

Therefore, we have

$$
\Delta u_{n,z} = \Delta\Phi
\tag{26}
$$

Eq. 26 suggests that the subgame of the follower follows an exact potential game and there exists at least one NE.

According to the positive linear correlation between the potential function and the total loss of the system, the optimal pure strategy NE is the global optimal solution of the problem. This completes the proof. ∎